\documentclass[preprint,review,12pt]{elsarticle}
\usepackage[margin=2.5cm]{geometry}

\usepackage{amssymb}
\usepackage{amsmath}
\usepackage{amsthm}
\usepackage{soul}
\usepackage{color}
\usepackage{float}

\makeatletter
\def\ps@pprintTitle{%
  \let\@oddhead\@empty
  \let\@evenhead\@empty
  \let\@oddfoot\@empty
  \let\@evenfoot\@oddfoot
}
\makeatother

\begin{document}
\begin{frontmatter}

\title{Enhanced Graphene–Water Thermal Transport via Edge Functionalization without Compromising In-Plane Thermal Conductivity}
\author[inst1]{John Crosby}
\author[inst1]{Haoran Cui}\cortext[mycorrespondingauthor]{Corresponding author}\ead{hcui@unr.edu}
\author[inst1]{Mehrab Lotfpour}            
\author[inst1]{Yan Wang}\cortext[mycorrespondingauthor]{Corresponding author}\ead{yanwang@unr.edu}
\author[inst1]{Lei Cao}\cortext[mycorrespondingauthor]{Corresponding author}\ead{leicao@unr.edu}

\affiliation{organization={Department of Mechanical Engineering, University of Nevada, Reno},
            city={Reno},
            postcode={89557}, 
            state={NV},
            country={USA}}

\begin{abstract}
Interfacial thermal transport between graphene and water plays a critical role in a wide range of thermal and energy applications. Although chemical functionalization can significantly enhance graphene–water interfacial thermal conductance, it often degrades graphene’s intrinsic in-plane phonon transport. In this work, we perform a systematic deep neural network molecular dynamics study comparing edge-functionalized graphene nanoribbons with surface-functionalized graphene in aqueous environments. We demonstrate that functionalizing only 10\% of the ribbon edges with hydroxyl groups increases the graphene–water interfacial thermal conductance by more than eightfold, primarily due to strengthened interfacial interactions and improved wettability at the edges. In contrast to basal-plane oxidation, edge functionalization largely preserves in-plane thermal conductivity. Importantly, hydroxyl edge groups exert competing effects on phonon transport: they introduce additional boundary scattering that suppresses heat conduction, while simultaneously passivating dangling bonds at bare edges, thereby reducing phonon localization and edge-induced scattering. This competition leads to a non-monotonic dependence of in-plane thermal conductivity on edge functionalization ratio. These results establish edge functionalization as an effective strategy for enhancing graphene–water interfacial thermal transport without sacrificing intrinsic phonon transport properties.
\end{abstract}

\end{frontmatter}

\section{Introduction}

Graphene, a two-dimensional allotrope of carbon composed of sp$^2$-bonded atoms arranged in a honeycomb lattice, exhibits extraordinary thermal conductivity ($\kappa$), exceeding 1000~W/m$\cdot$K in its pristine single-layer form \cite{balandin2008superior,xu2014length,wang2014two}. This exceptional thermal property, together with its large specific surface area and mechanical flexibility, makes graphene an attractive material for a wide range of thermal management applications \cite{shtein2015ACSAMI,cho2016Carbon,guo2021ACSnano,cui2024TwoD}. In particular, graphene has been extensively utilized in aqueous environments, such as solar vapor generation systems \cite{yang2017nano,wu2019multifunctional,fu2018oxygen} and photon-based (laser or microwave) processing of graphene materials in solution \cite{C6NR01148A,WANG2020880}. In these applications, efficient interfacial heat dissipation from graphene nanostructures to the surrounding water critically determines device performance or processing outcome.

Despite its technological relevance, interfacial thermal transport between graphene-based nanostructures and water remains insufficiently understood. Only a limited number of atomistic modeling studies \cite{alexeev2015NL,li2022IJHMT,Cui2025Carbon} and even fewer experimental investigations \cite{huxtable2003NatureMat, casto2024Carbon} have quantified the interfacial thermal conductance ($G$) at graphene–water or carbon nanotube-water interfaces. Direct experimental measurements, using pump–probe techniques, for example, are challenged by strong optical absorption over a broad wavelength range and the intrinsically low $\kappa$ of bulk water, both of which significantly reduce measurement sensitivity and accuracy.

From a computational perspective, atomistic modeling of graphene–water interfacial heat transfer has historically been hindered by the lack of high-fidelity interatomic potentials capable of accurately describing graphene (particularly oxidized graphene), water, and their cross-interactions within a unified framework. Nevertheless, several simulations have shown that $G$ between pristine graphene and water is limited by weak van der Waals interactions, typically ranging from 20 to 36~MW/m$^2\cdot$K \cite{alexeev2015NL,li2022IJHMT,Cui2025Carbon}. This relatively low $G$ presents a bottleneck for efficient heat dissipation in aqueous environments.

To overcome this limitation, chemical functionalization has been proposed as an effective strategy to enhance phonon coupling across the interface and increase $G$. Functional groups can modify the local vibrational density of states and introduce stronger interfacial interactions, such as hydrogen bonding or even partial covalent bonding, between graphene and water molecules. In our recent work \cite{Cui2025Carbon}, we demonstrated, using molecular dynamics (MD) simulations with a high-accuracy deep neural network (DNN) interatomic potential, that surface functionalization of graphene with hydroxyl and epoxide groups can enhance $G$ between graphene and water by nearly an order of magnitude. Such enhancement has important implications for applications requiring either efficient heat dissipation or, conversely, controlled thermal insulation between graphene and aqueous media.

Functionalization strategies can generally be classified into two categories: surface functionalization and edge functionalization \cite{kuila2012chemical,georgakilas2012functionalization,bellunato2016chemistry,dai2013functionalization}. Surface-functionalized graphene (SFG) involves covalent or noncovalent attachment of functional groups—such as hydroxyl (–OH), carboxyl (–COOH), or epoxide (–O–)—to the basal plane. This approach is commonly achieved via oxidative treatments, such as Hummers' method \cite{chen2013improved,shahriary2014graphene,zaaba2017synthesis}, to produce graphene oxide (GO), followed by chemical or thermal reduction \cite{chen2010preparation,pei2012reduction,de2017chemical,agarwal2021strategies}. While surface functionalization substantially enhances hydrophilicity and interfacial adhesion with water \cite{wang2009synthesis,Cui2025Carbon}, it disrupts the $\pi$-conjugated network, introduces sp$^3$-hybridized defects, and significantly reduces graphene’s intrinsic in-plane $\kappa$ \cite{chien2012influence,mu2014thermal,chen2020effect,Cui2025Carbon}. The symmetry breaking associated with sp$^3$ bonding increases anharmonic scattering of flexural phonons, leading to pronounced degradation of thermal transport.

In contrast, edge-functionalized graphene (EFG) preserves the pristine basal plane while selectively introducing functional groups at the ribbon edges. This strategy largely maintains the high in-plane $\kappa$ of graphene \cite{guo2009thermal,Wang2012APL,bae2013ballistic}, making it attractive for applications requiring efficient lateral heat spreading. Edge functionalization can be realized through plasma etching, chemical cutting, or nanotube unzipping, followed by targeted edge chemistry. Chemically, EFG introduces localized reactive sites capable of forming strong interactions with adjacent media without significantly perturbing phonon transport within the basal plane.

Despite these advances, a systematic comparison between SFG and EFG with respect to graphene–water interfacial thermal transport remains lacking. Such a comparison is essential, as the practical design of graphene-based thermal interfaces requires balancing two competing objectives: maximizing $G$ while preserving the intrinsic in-plane $\kappa$. In this work, we address this gap through large-scale MD simulations enabled by a DNN interatomic potential trained on \emph{ab initio} data. We investigate interfacial heat transfer between graphene and water, with particular emphasis on the distinct roles of surface and edge functionalization. Our simulations mimic transient laser heating of graphene immersed in water and quantify $G$ at both SFG–water and EFG–water interfaces. Furthermore, we analyze the structural configurations and interfacial interaction characteristics that govern the observed differences in $G$. This study establishes a mechanistic understanding of functionalization-dependent interfacial thermal transport and provides quantitative design guidelines for optimizing graphene-based thermal systems operating in liquid environments.

\section{Methodology}

\subsection{Development of Deep Neural Network Interatomic Potential}

The DNN interatomic potential developed in this work is based on the DeePMD framework \cite{wang2018deepmd, cui2025CMS} , trained using a comprehensive dataset generated from ab initio molecular dynamics (AIMD) simulations. While our previous work \cite{Cui2025Carbon, nasiri2026deep} involved the construction of a DNN potential for the GO–water system, it was limited to configurations featuring SFGs. To ensure reliable modeling of EFGs, i.e., graphene nanoribbon oxides (GNROs) in the current study, we extended the original dataset by performing additional AIMD simulations that include a range of edge terminations.

Specifically, we generated new AIMD data for graphene nanoribbons (GNRs) with carboxyl, hydroxyl, and hydrogen edge groups, as well as GNRs with bare edges containing dangling bonds. Both dry and water-immersed configurations were considered to capture a broad spectrum of atomic force environments relevant to thermal transport phenomena in GNR–water systems. Simulations were conducted under varying temperatures and pressures to enhance the diversity and transferability of the training set. The complete list of AIMD configurations and corresponding thermodynamic conditions is summarized in Table~\ref{tab:table1}.

\begin{table}
    \centering
    \begin{tabular}{lllll}
    Structure&Ensemble&Temperature (K)&Pressure (GPa)&Number of Data\\
    \hline
    Graphene & NVT & 200-1,000& N/A & 25,931\\
     \hline
    Graphene-water & NPT & 300-1,000& 0-0.1& 10,000\\
     \hline
    \begin{tabular}{l}GO(5-70\%O/C)\\of 8 configurations\end{tabular} & NVT & 300-1,000& N/A & 75,549
\\
     \hline
    \begin{tabular}{l}GO(5-70\%O/C)-water\\of 8 configurations\end{tabular} & NVT& 300-800& N/A& 28,022\\
 \hline
 \begin{tabular}{l}GO(5-70\%O/C)-water\\of 8 configurations\end{tabular} & NPT& 300-1,000& 0-1.0&7,596\\
     \hline
    \begin{tabular}{l}5\%O/C CNTO-water\\Diameter=$0.55$ nm\end{tabular} & NPT & 900 & 0 & 5,000\\
 \hline
 Dry GNRO/GO flakes& NVT& 300-2,500& N/A&4891\\

\hline
 GNRO/GO flakes in water& NPT& 300-700& 0&6684\\

 \hline
    \end{tabular}
    \caption{List of the structures and thermodynamic conditions for AIMD data used to train the DNN potential for this work.}
    \label{tab:table1}
\end{table}

All AIMD simulations were performed using the Vienna \textit{ab initio} Simulation Package (VASP) \cite{kresse1993VASP, kresse1996CMS, kresse1999PRB, hafner2008Vasp}. The projector augmented-wave (PAW) method was used to describe the electron–ion interactions, while the exchange-correlation effects were treated using the Perdew–Burke–Ernzerhof (PBE) functional within the generalized gradient approximation (GGA). A plane-wave energy cutoff of 600~eV and a Gaussian smearing width of 0.05~eV were applied. The electronic self-consistency loop was converged to $10^{-6}$~eV. To account for dispersion forces, van der Waals interactions were included using the Becke–Johnson damping scheme. $\Gamma$-point sampling was employed due to the large supercell dimensions ($>$15~\AA~in all directions). For each simulation, time-resolved atomic trajectories, forces, energies, and cell vectors were collected to serve as training data for the DNN potential.

The deep potential model was trained using the DeePMD-kit package \cite{wang2018deepmd}. The descriptor construction was based on full relative atomic coordinates with a local environment cutoff of 6.0~\AA~and a smoothing region of 0.5~\AA. The embedding network consists of three hidden layers with 25, 50, and 100 neurons, respectively. The fitting network comprises three hidden layers, each with 240 neurons. The training process was conducted with an initial learning rate of 0.001. A decay step schedule was applied every 16 million steps with an exponential decay factor of $3.51 \times 10^{-8}$. These hyperparameters were selected based on best practices documented in the DeePMD-kit reference guides and validated by convergence behavior observed during training.

\subsection{Equilibrium molecular dynamics}
Equilibrium molecular dynamics (EMD) simulations were performed in LAMMPS \cite{thompson2022lammps} to determine the bulk-limit in-plane $\kappa$ along the length ($x$) direction of GNROs of varying edge functionalization ratios and widths. The simulation domain was periodic only in the $x$ direction, while the finite width ($y$) and cross-plane ($z$) directions were treated with free boundaries to eliminate interactions between periodic images in these directions. For cases with edge functionalization ratio as the independent variable, the GNRO's dimensions were fixed at 8.34 nm along the $x$ direction and 2.32 nm along the $y$ direction. For cases with width as the independent variable, the width of the GNRO was varied from 2.32 nm to 36.6 nm. Beyond this range, structural instability arising from excessive width rendered $\kappa$ measurements meaningless. A convergence study confirmed that 8.34 nm was sufficiently long to produce converged $\kappa$ (see Supplementary Materials). The timestep for all EMD simulations in this study was 0.5 fs. The entire simulation was conducted under the canonical (NVT) ensemble to prevent undesirable crumpling of the GNRO along the periodic direction during isobaric relaxation. The system was first heated from 5 K to 300 K over 100 ps, followed by a 250 ps equilibration at 300 K. Subsequently, the GNRO was allowed to freely evolve at 300 K for 1 ns, during which heat current fluctuations were recorded along the periodic direction. $\kappa$ was extracted using the Green-Kubo formalism \cite{volz2000PRB,cui2024PRB}, derived from the fluctuation-dissipation theorem, which relates random heat fluctuations at equilibrium to macroscopic non-equilibrium thermal dissipation. $\kappa$ of GNRO is therefore calculated as:
\begin{equation}
    \kappa=\frac{1}{\Omega k_b T^2}\int_{0}^{\tau_m}\left<J(\tau)J(0)\right>d\tau,
    \label{greenkubo}
\end{equation}
where $\Omega$ is the system volume, $k_b$ is the Boltzmann constant, and $T$ is the equilibrium temperature. The term $\left<J(\tau)J(0)\right>$ denotes the heat current autocorrelation function, representing the time-averaged correlation of the instantaneous initial heat flux ($J(0)$) and that at time $\tau$ ($J(\tau)$).

\subsection{Transient molecular dynamics and lumped capacitance model}

To quantify the effect of edge functionalization on interfacial thermal transport between GNRO and water, we performed transient MD simulations that mimic rapid laser heating of the GNRO followed by heat dissipation into the surrounding aqueous environment. In each simulation, the combined GNRO–water system was first equilibrated using a Nosé–Hoover thermostat and barostat~\cite{Nose1984JCP,Hoover1985PRA} in two successive NPT stages. The timestep for all transient MD simulations was 0.1~fs.

\begin{figure} 
\centering \includegraphics[width=0.5\textwidth, keepaspectratio]{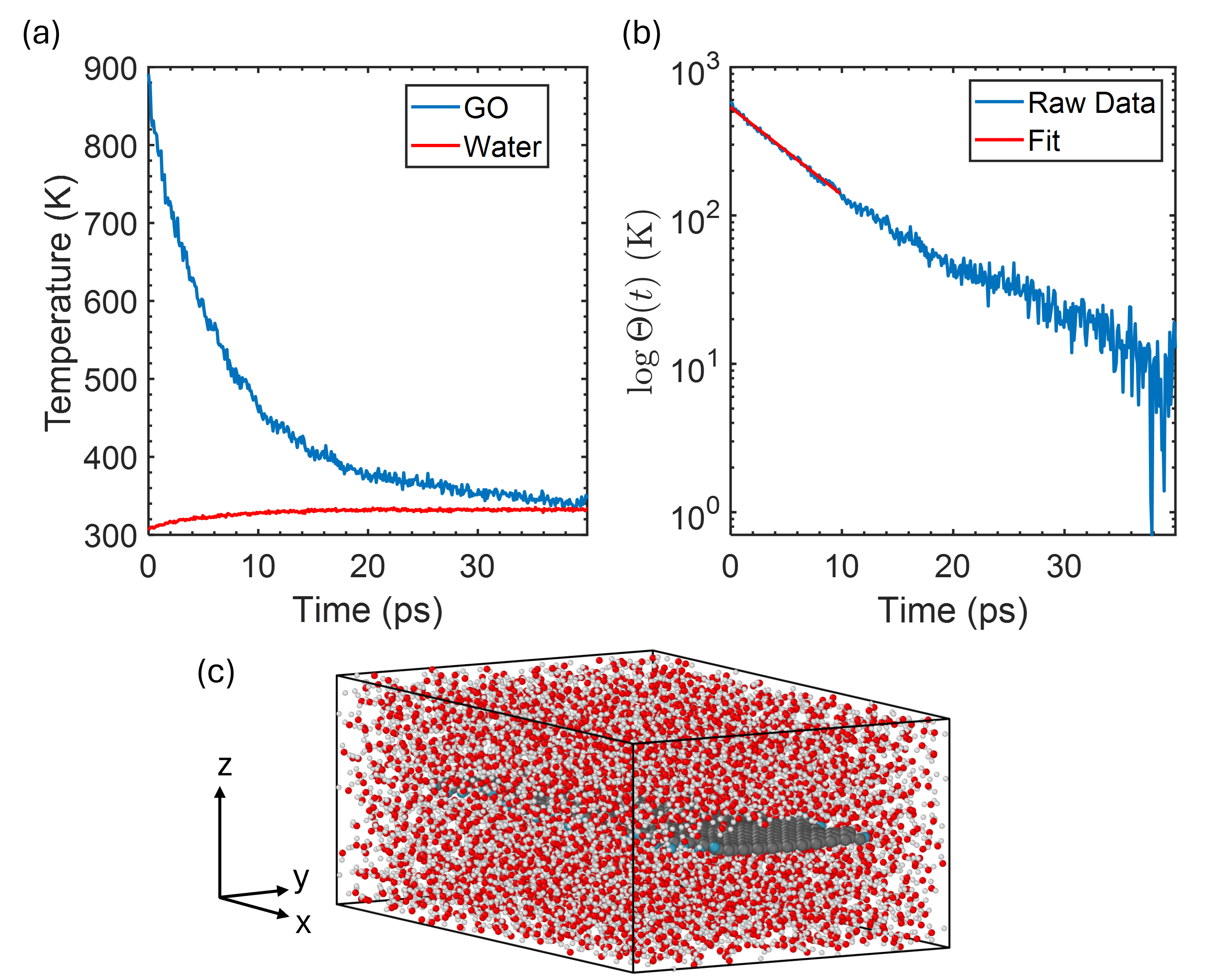} \caption{Time dependent (a) temperature and (b) temperature difference decay curves. $G$ is extracted from the linear fit shown in panel (b). (c) A representative structure for transient MD simulations used in this study. The box is periodic in the $x$ direction and padded with enough water in the $y$ direction to prevent interactions across the periodic boundary. Grey atoms represent carbon atoms, blue atoms represent hydroxyl functional groups, while red and white atoms represent oxygen and hydrogen of water, respectively.} 
\label{fig:transient} 
\end{figure}

In the first stage, the system was heated from 5~K to 300~K over 50~ps using a linear temperature ramp, followed by isothermal equilibration at 300~K for an additional 50~ps. The GNRO temperature was then instantaneously elevated to 900~K via atomic velocity rescaling implemented in LAMMPS and maintained for 1~ps. Subsequently, the entire system was allowed to evolve freely under the microcanonical (NVE) ensemble for up to 100~ps to reach thermal equilibrium (Fig.~\ref{fig:transient}a).

Interfacial thermal conductance $G$ between the GNRO and water was determined following the procedure described in our previous work~\cite{Cui2025Carbon}. Specifically, $G$ was extracted by fitting an analytical solution to the transient temperature decay profiles. This analytical solution is derived from the lumped capacitance approximation, which yields two coupled differential equations:
\begin{equation}
    \frac{A_{sur}G(T_s-T_w)}{C_sV_s}=\frac{dT_s}{dt},
    \label{lumped_solid}
\end{equation}
and
\begin{equation}
    \frac{-A_{sur}G(T_s-T_w)}{C_wV_w}=\frac{dT_w}{dt},
    \label{lumped_liquid}
\end{equation}
where $A_{sur}$, $C$, $V$, $T$, and $t$ denote the interfacial area, specific heat capacity, volume, temperature, and time, respectively. $A_{sur}$ is calculated as twice the total basal-plane area of the GNRO, since both sides are in contact with water. Each carbon atom contributes $\frac{3\sqrt{3}a^2}{4}\,\text{\AA}^2$ to the total area, where $a=1.41\,\text{\AA}$ is the C–C bond length used in this study. Atoms belonging to functional groups are excluded to ensure consistent comparison across different oxidation ratios. Subscripts $s$ and $w$ denote the solid (which is GNRO in this work) and water, respectively.

Subtracting Eq.~\eqref{lumped_liquid} from Eq.~\eqref{lumped_solid} yields
\begin{equation}
    \frac{d\Theta}{dt}=-\frac{\Theta}{\tau_t},
    \label{dthetadt}
\end{equation}
where we have defined the temperature difference between the GNRO and water as
\begin{equation}
    \Theta = T_s - T_w,
    \label{thetadiff}
\end{equation}
and defined the characteristic relaxation time as
\begin{equation}
    \frac{1}{\tau_t}=A_{sur}G\left(\frac{1}{C_sV_s}+\frac{1}{C_wV_w}\right).
    \label{taut}
\end{equation}

The solution to Eq.~\ref{dthetadt} is
\begin{equation}
    \Theta(t)=\Theta(0)\exp\left(-\frac{t}{\tau_t}\right).
    \label{eqn:final}
\end{equation}

The relaxation time $\tau_t$, and thus $G$ (via Eq.~\ref{taut}), is obtained by least-squares fitting of Eq.~\ref{eqn:final} to the transient MD temperature decay curves (Fig.~\ref{fig:transient}b).

Transient simulations are conducted for both varying edge oxidation ratios and varying widths. A sample transient simulation domain is shown in Fig.~\ref{fig:transient}c. The simulation box is periodic in the $x$ direction, while sufficient water molecules are added in the $y$ and $z$ directions to prevent interactions of the GNRO with its periodic image in these directions. In the cases where width is the independent variable, extra water padding is added to the simulation box in the finite width ($y$) direction to maintain a consistent GNRO weight percent of approximately 10.5\% in all simulations, ensuring a fair comparison across varying GNRO widths. 

\subsection{Cross-sectional heat flux distribution}
\begin{figure}
    \centering
\includegraphics[width=0.5\textwidth, keepaspectratio]{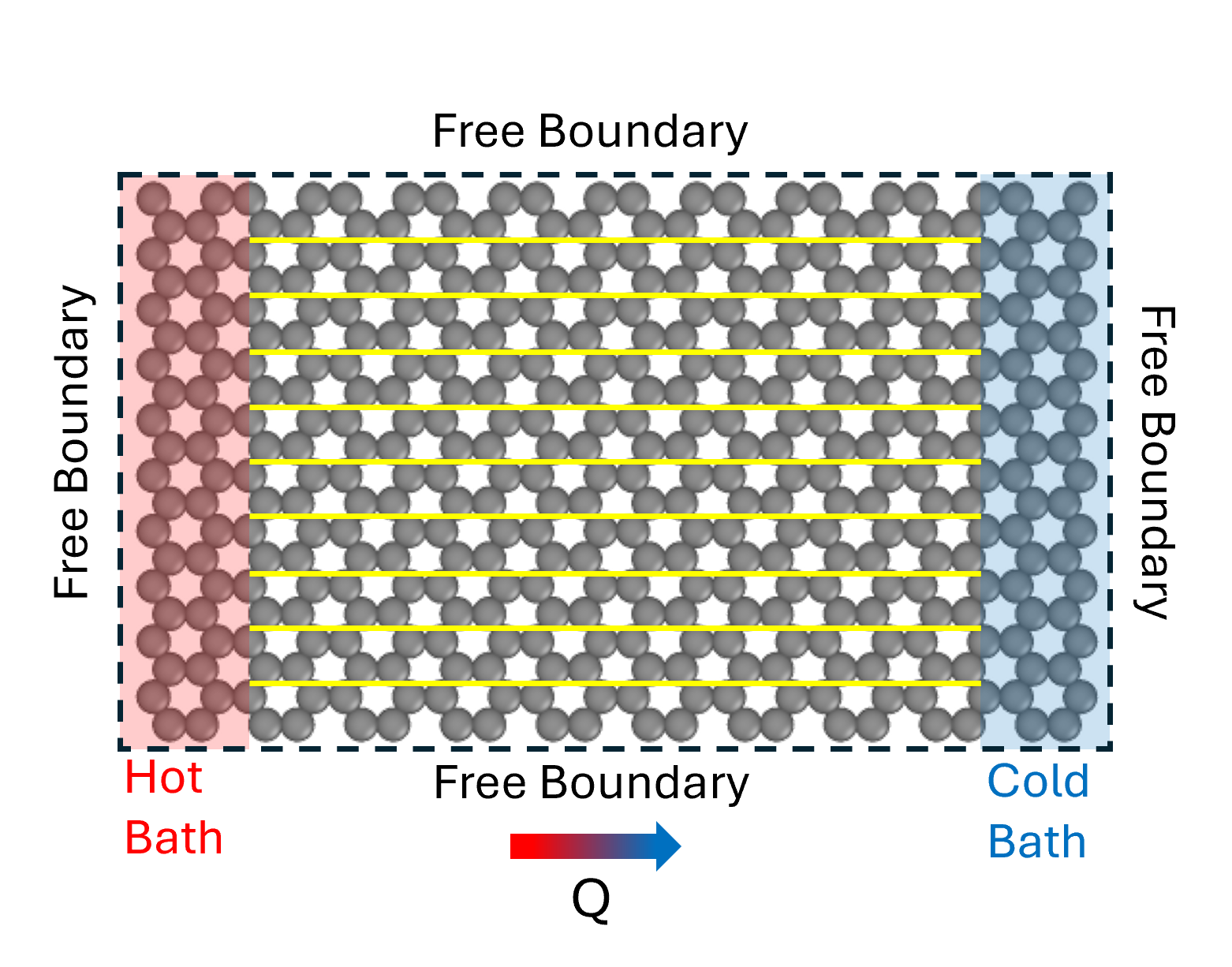}
    \caption{Diagram of cross-sectional heat flux distribution simulation configuration.}
    \label{fig:CrossSectionalSetup}
\end{figure}

To further elucidate the effect of edge functionalization on phonon transport throughout the GNRO, we conducted cross-sectional heat flux distribution calculations following a non-equilibrium process detailed in a previous work \cite{Wang2012APL}. The GNRO is divided into a hot bath, cold bath, and device region. Free boundary conditions are imposed on each direction of the simulation box (see Fig. \ref{fig:CrossSectionalSetup}) to prevent heat transfer between periodic images of the GNRO edges and surface, as well as direct heat transfer between the hot and cold baths. The supercell is first relaxed to ensure all atoms are in their equilibrium positions, then atoms are assigned to specific columns. Langevin thermostats are used to maintain the hot bath and cold bath at 330 K and 270 K respectively, which imposes a thermal gradient on the device region of the GNRO. The device region is divided into ten equally-sized columns along the width direction $y$, and the heat flux along the length ($x$) direction is calculated both per-column ($J_{x,i}$) and for the entire GNRO ($J_{x,all}$). Since the heat current is inherently noisy due to the statistical nature of MD simulations, both heat currents are integrated with respect to time, yielding cumulative heat fluxes on an overall and per-column basis. The value we report is therefore $\alpha_{i} = \frac{Q_{x,i}}{Q_{all}}$, which describes the fraction of total heat carried by the $i$th column. 

\section{Results and Discussions}
\subsection{The effect of edge functionalization on thermal transport in GNRO\label{sec:k}}
While prior theoretical studies \cite{lan2009edge,xie2011thermal,aksamija2011lattice,Wang2012APL} have explored thermal transport in GNRs with bare or hydrogen-passivated edges, a systematic investigation of edge functionalization by oxygen-containing groups remains lacking. In this context, the application of DNN potential-based MD simulations, enabled by the first ab initio–derived DNN potential for such systems, provides a robust framework to examine how the $\kappa$ of GNRs is influenced by oxygen-based edge functionalization. Such functional groups represent the most prevalent form of edge chemistry in GNRs and graphene flakes encountered in experimental studies and practical applications, underscoring the relevance of this investigation.

\begin{figure}
    \centering
\includegraphics[width=1\textwidth, keepaspectratio]{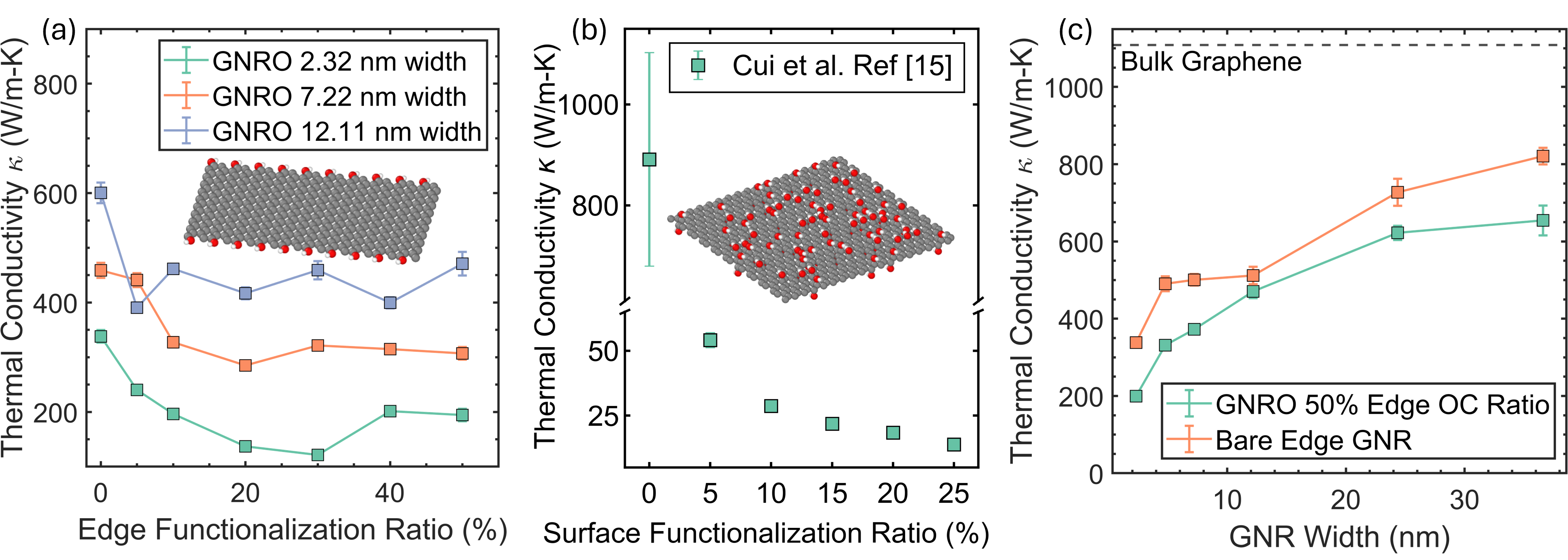}
    \caption{ $\kappa$ dependence on (a) OH edge functionalization ratio of GNRO of varying widths, (b) on OH surface functionalization ratio for GO flake, (c) on nanoribbon width for both pristine and 50\% edge-functionalized nanoribbon.}
    \label{fig:kGNRO}
\end{figure}

Figure~\ref{fig:kGNRO}a illustrates the dependence of $\kappa$ on the edge functionalization ratio, which is defined as the ratio of oxidized edge carbon atoms to the total number of edge carbon atoms, for GNROs with widths of 2.32, 7.22, and 12.11~nm. For all three widths, $\kappa$ initially decreases with increasing edge functionalization ratio.

For the narrowest GNRO (2.32~nm), this decrease is relatively gradual and reaches a minimum at functionalization ratios of approximately 20–30\%, beyond which $\kappa$ increases with further oxidation. In contrast, this non-monotonic behavior becomes progressively weaker for wider GNROs. Specifically, the 7.22-nm-wide GNRO exhibits only a slight reduction in $\kappa$ when the functionalization ratio increases from 0\% (bare edges) to 20\%, while the 12.11-nm-wide GNRO shows minimal dependence of $\kappa$ on the degree of edge functionalization beyond 5\%.

The observed non-monotonic dependence of $\kappa$ on edge functionalization originates from the competition between two opposing mechanisms. At low functionalization ratios, the introduction of oxygen-containing functional groups increases structural disorder at the GNRO edges, enhancing phonon scattering and phonon localization, which leads to a reduction in $\kappa$. As edge oxidation increases further, however, functional groups progressively saturate dangling bonds at the ribbon edges, partially alleviating edge-induced disorder and suppressing phonon scattering and localization. This edge passivation effect counterbalances the disorder-induced scattering at higher oxidation levels, resulting in a saturation of $\kappa$ for wider GNROs and a moderate recovery of $\kappa$ for narrower GNROs. With increasing GNRO width, the relative contribution of edge effects to overall thermal transport diminishes, thereby reducing the sensitivity of $\kappa$ to variations in the edge functionalization ratio.

GNRs can be functionalized at the edges, on the basal plane, or simultaneously at both locations. It is therefore instructive to compare the relative impacts of edge and surface functionalization on thermal transport in graphene-based structures. Figure~\ref{fig:kGNRO}b presents $\kappa$ of surface-functionalized GO as a function of the surface functionalization ratio, as predicted in our recent study \cite{Cui2025Carbon}. Clearly, $\kappa$ decreases abruptly from nearly 1000~W\,m$^{-1}$\,K$^{-1}$ for non-functionalized GO (0\% functionalization) to approximately 50~W\,m$^{-1}$\,K$^{-1}$ at a surface functionalization ratio of 5\%, corresponding to a reduction of more than 90\%. This reduction is substantially larger than that induced by edge functionalization, which decreases $\kappa$ of GNRs with widths of 2.32, 7.22, and 12.11~nm by only 29\%, 4\%, and 34\%, respectively, at the same 5\% functionalization ratio. This stark contrast clearly demonstrates that surface functionalization has a much stronger impact on phonon transport than edge functionalization in graphene.

The pronounced effect of surface functionalization can be attributed to the conversion of \textit{sp}$^2$ bonds to \textit{sp}$^3$ bonds upon functionalization of basal-plane carbon atoms, which transforms the originally flat graphene lattice into a corrugated, non-planar structure. The exceptionally high intrinsic $\kappa$ of pristine graphene is widely understood to arise from the low scattering rates of flexural phonon modes, whose anharmonic three-phonon scattering phase space is severely constrained by mirror symmetry of the atomically flat graphene layer \cite{lindsay2010flexural}. When this symmetry is broken by the formation of \textit{sp}$^3$ bonds associated with surface functionalization, the anharmonic scattering rates of flexural phonons increase dramatically, leading to a substantial suppression of thermal transport. In contrast, edge functionalization has a less significant effect on the flatness of the graphene layer, thus it mostly affects phonon transport by serving as a boundary scattering or phonon localization mechanism. 

Figure \ref{fig:kGNRO}c shows the dependence of $\kappa$ on GNRO width. The persistent increase in $\kappa$ with ribbon width in both cases arises from graphene's intrinsically large phonon mean free paths, which are orders of magnitude larger than typical GNRO widths. The dominant heat carriers in graphene, primarily acoustic phonon modes, exhibit mean free paths that can extend into the micrometer scale and are significantly suppressed by edge scattering in finite-width ribbons. As the width of the ribbons increases, boundary scattering and edge localization has a diminishing suppressive effect on phonon transport, and allows a larger fraction of acoustic phonons to contribute to thermal transport, resulting in increased $\kappa$. The lack of saturation, even in ribbons up to 36.6 nm wide, reflects this mechanism; saturation of $\kappa$ will not be observed until the width of the ribbon approaches the mean free path of graphene's primary heat carriers.

\begin{figure}
    \centering
\includegraphics[width=1\textwidth, keepaspectratio]{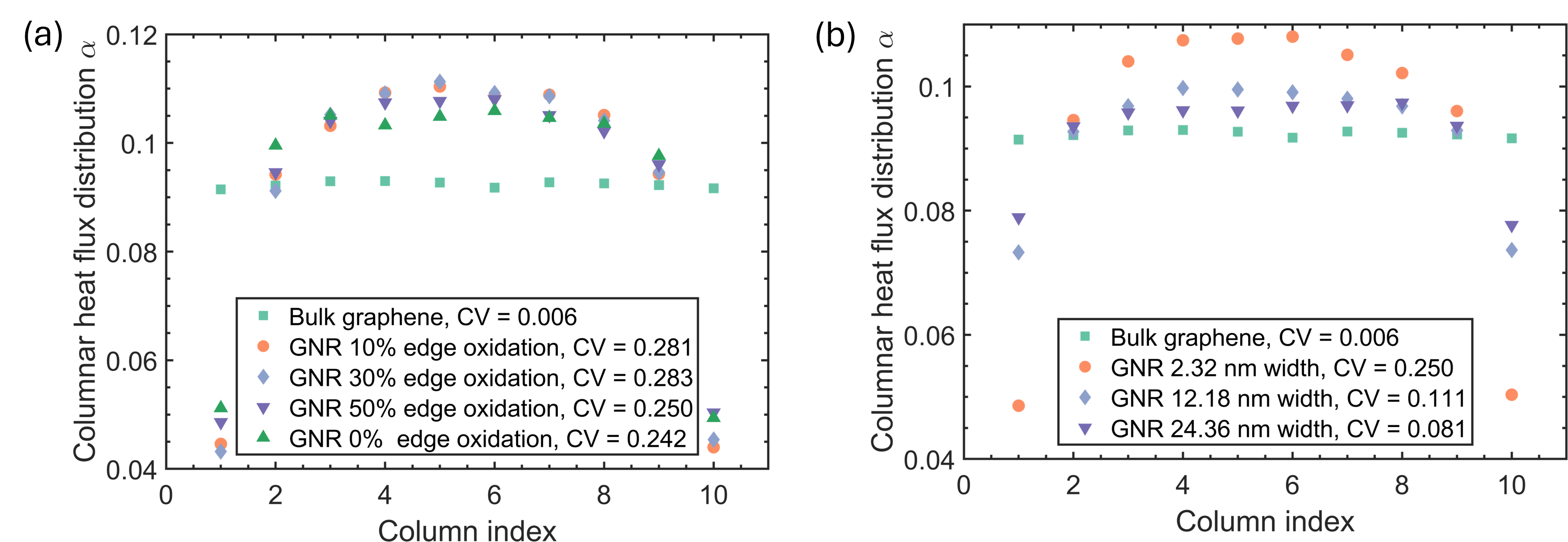}
    \caption{Cross-sectional heat flux distribution, $\alpha$, for (a) a fixed GNRO width (2.32 nm) with varying edge functionalization ratios, and (b) a fixed edge functionalization ratio (50\%) with varying GNRO widths. The CV values reported in the legend are defined as the ratio of the standard deviation of $\alpha$ to its mean value, providing a quantitative measure of the dispersion of $\alpha$.}
    \label{fig:CrossSectionalHeatFlux}
\end{figure}

To elucidate the influence of edge chemistry, including bare edges and edges functionalized at different ratios, on thermal transport in GNROs, we analyze the columnar distribution of heat flux, denoted by $\alpha$, as shown in Fig.~\ref{fig:CrossSectionalHeatFlux}. This metric directly reflects the spatial distribution of heat flow as a function of distance from the GNRO edges.

As shown in Fig.~\ref{fig:CrossSectionalHeatFlux}a, bulk graphene, which is free of edges, exhibits a nearly uniform columnar distribution of $\alpha$, as expected. To quantitatively assess the degree of non-uniformity, we define the coefficient of variance (CV) as the ratio of the standard deviation of $\alpha$ to its mean value. For bulk graphene, the CV is as low as 0.006. This small but finite value originates from intrinsic statistical fluctuations in MD simulations, which lead to instantaneous deviations from perfectly uniform heat flow even in homogeneous materials.

In contrast to bulk graphene, GNROs with intermediate edge functionalization ratios exhibit pronounced suppression of heat flux near the edges relative to the central region, as shown in Fig.~\ref{fig:CrossSectionalHeatFlux}a. This behavior leads to significantly larger CV values of 0.281 and 0.283 for edge functionalization ratios of 10\% and 30\%, respectively. When the edge functionalization ratio is further increased to 50\%, however, the GNRO exhibits reduced variance in the columnar heat flux distribution, with a lower CV of 0.250. These trends are consistent with the competing mechanisms inferred from the $\kappa$ data in Fig.~\ref{fig:kGNRO}a. At low edge functionalization ratios, the introduction of functional groups enhances boundary (edge) phonon scattering and phonon localization, thereby suppressing thermal transport, particularly in edge-adjacent regions. As the edge functionalization ratio increases, additional functional groups progressively saturate dangling bonds, partially mitigating boundary scattering and enabling increased heat transport near the edges. This interpretation is consistent with the higher $\kappa$ observed for the 50\%-edge-functionalized GNRO compared to the bare-edge (0\% functionalized) GNR in Fig.~\ref{fig:kGNRO}a.

Figure~\ref{fig:CrossSectionalHeatFlux}b presents the columnar heat flux distributions for GNROs with a 50\% edge functionalization ratio and different ribbon widths. As shown, narrower GNROs exhibit substantially higher heat flux in the central region than near the edges, clearly demonstrating the suppressive effect of edges on phonon transport. Quantitatively, the 2.32-nm-wide GNRO exhibits a relatively large CV of 0.250, whereas the CV decreases to 0.111 and 0.081 for GNROs with widths of 12.18~nm and 24.36~nm, respectively.

Figure S2 in the Supplementary Materials presents $\alpha$ for bare-edge GNRs, exhibiting characteristics similar to those of the 50\%-edge-functionalized GNROs shown in Fig.~\ref{fig:CrossSectionalHeatFlux}b. However, for a given ribbon width, the variance, quantified by CV, is smaller than that of the corresponding 50\%-edge-functionalized GNROs, indicating weaker edge-induced phonon scattering in the absence of functional groups. In both cases, the variance decreases with increasing ribbon width, leading to flatter and more spatially uniform heat flux distributions in wider nanoribbons. This behavior is consistent with the $\kappa$ results shown in Fig.~\ref{fig:kGNRO}c, where wider GNRs exhibit higher in-plane $\kappa$.

\subsection{Structure of GNRO-water interface}

Before analyzing heat transfer across the GNRO–water interface, it is essential to first examine the equilibrated interfacial structures. In the GNRO–water systems studied here, four distinct types of interfaces can be identified: (1) the interface between a bare graphene or GNR basal plane and water, where water molecules interact directly with carbon atoms; (2) the interface between a functionalized basal plane and water, where water molecules interact directly with surface functional groups and only indirectly with the underlying carbon lattice; (3) the interface between a bare GNR edge and water molecules; and (4) the interface between edge functional groups and water molecules. These interfacial configurations give rise to markedly different local bonding environments and interaction mechanisms, which are expected to play a critical role in interfacial thermal transport. 

\begin{figure}
    \centering
\includegraphics[width=1.0\textwidth, keepaspectratio]{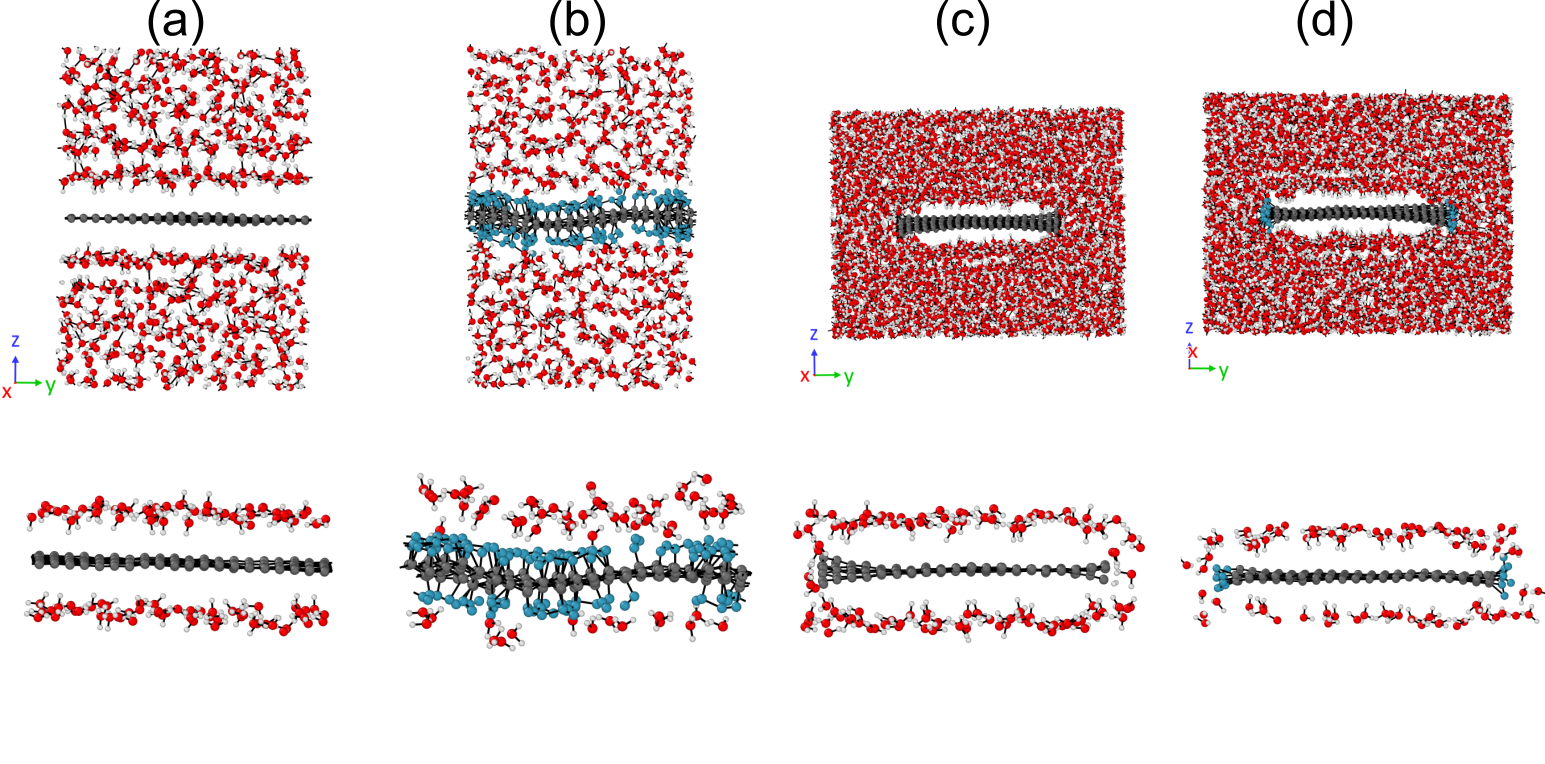}
    \caption{Atomic configurations of water molecules interacting with (a) the bare GNR basal plane, (b) the functionalized basal plane, (c) the bare GNR edge, and (d) the functionalized GNR edge.}
    \label{fig:structures}
\end{figure}

Figure \ref{fig:structures} presents representative equilibrated structures corresponding to these four interfacial scenarios, as obtained from our DNN-MD simulations after equilibriation at 300 K and atomosphere pressure in the isothermal–isobaric ensemble for 500 ps. As shown in Fig. \ref{fig:structures}a, there is a significant gap between water molecules and bare GNR basal plane, which is indeed due to the rather weak interaction between bare graphene surface and water. This behavior is consistent with the hydrophobic nature of pristine graphene. In contrast, the functionalized basal plane, as displayed in Fig. \ref{fig:structures}b, exhibits significantly reduced spacing between GO layer (its functional groups) and water, indicating enhanced attraction of water molecules to the GO surface. A further examine of interface structure reveals notable hydrogen bonding between water molecules and hydroxyl groups at the GO surface. 

Figures~\ref{fig:structures}c and d show the interactions between water molecules and the bare and functionalized GNR edges, respectively. While both cases exhibited bonding between water molecules and edge C–C sites, functionalization strengthened water adsorption at the edge. For the bare GNR edge, broken C–C bonds significantly improved bonding with water molecules and hydroxyl groups (Fig. \ref{fig:structures}c).

\subsection{Transient heating of GNROs in water}

\begin{figure}
    \centering
    \includegraphics[width=1\textwidth, keepaspectratio]{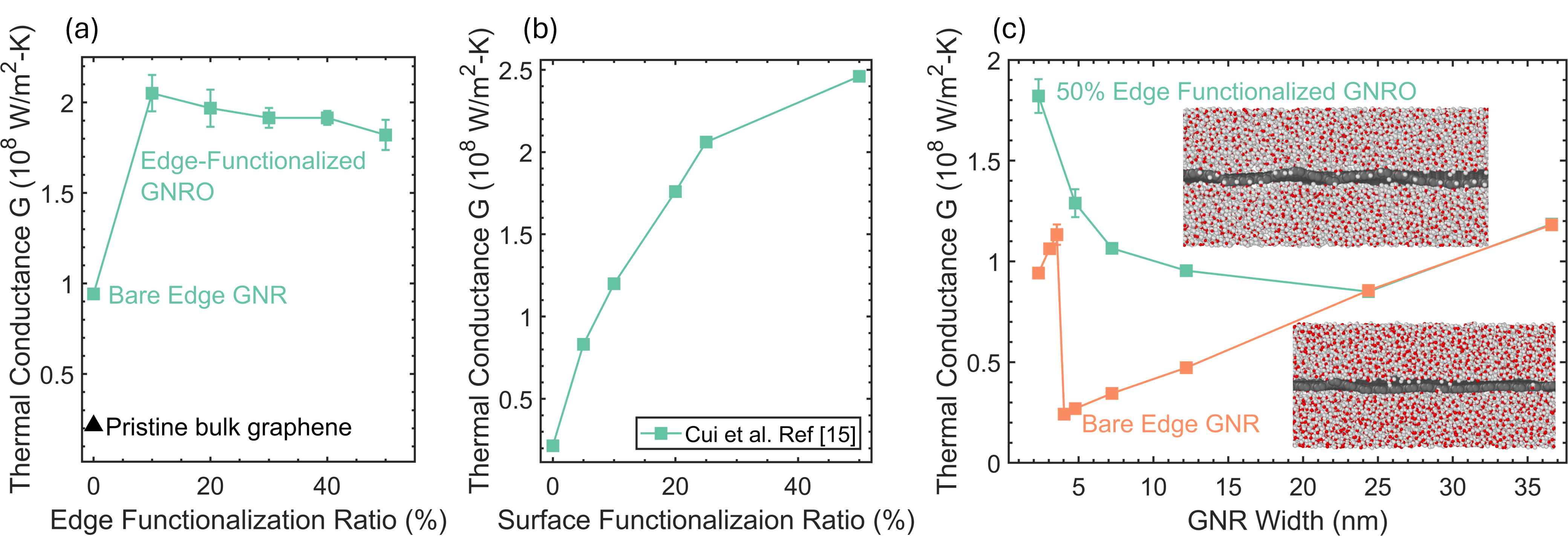}
    \caption{Dependence of $G$ on (a) edge functionalization ratio in GNRO, (b) surface functionalization ratio in GO flake, and (c) width in edge-functionalized GNROs. Insets in (c) show difference in passivation between 3.06 nm (top) and 3.55 nm (bottom) width GNR after relaxation, resulting in the non-monotonic trend of $G$ in initially bare-edge GNR.}
    \label{fig:G}
\end{figure}

Figure~\ref{fig:G}a presents the dependence of the interfacial thermal conductance between edge-functionalized GNROs and water, $G$, on the edge functionalization ratio. Notably, even the 0\%-functionalized GNR (i.e., bare-edge) exhibits a 338\% higher thermal conductance with water, $G=0.942 \times 10^{8}$ W/m$^{2}\cdot K$, than pristine bulk graphene ($G=0.215\times 10^{8}$ W/m$^{2}\cdot K$). This enhancement is consistent with the pronounced hydrophilicity of bare GNR edges, as shown in Fig.~\ref{fig:structures}c. 

Moreover, as the edge functionalization ratio increases from 0\% to 10\%, $G$ rises sharply from $G=0.942 \times 10^{8}$ W/m$^{2}\cdot K$ to $2.05 \times 10^{8}$ W/m$^{2}\cdot K$, corresponding to a further increase of more than 100\%. This substantial enhancement arises from improved interfacial wettability induced by edge functionalization: the polar hydroxyl groups at GNRO edges interact more strongly with surrounding water molecules than bare carbon edges, as illustrated in Figs.~\ref{fig:structures}c and d, thereby strengthening interfacial thermal coupling. 

However, when the edge functionalization ratio exceeds 10\%, $G$ approaches a plateau. At this level of functionalization, water molecules are already densely accumulated near the GNRO edges, approaching the maximum achievable interfacial packing density. Furthermore, as demonstrated in our recent study on surface-functionalized graphene oxide, adjacent functional groups may compete as parallel heat-transfer pathways to water, thereby limiting any additional increase in $G$ \cite{Cui2025Carbon}.

For comparison, Fig.~\ref{fig:G}b presents $G$ as a function of surface functionalization ratio, reproduced from our recent work \cite{Cui2025Carbon}. The enhancement in $G$ due to edge functionalization is clearly much weaker than that achieved by surface functionalization, which can increase $G$ by nearly an order of magnitude. Nevertheless, it is important to note that surface functionalization substantially degrades the in-plane $\kappa$ and alters other physical and chemical properties of graphene, whereas edge functionalization largely preserves these intrinsic characteristics. Consequently, edge functionalization may be a more favorable strategy for applications requiring enhanced interfacial heat transfer without compromising in-plane transport properties.

Figure~\ref{fig:G}c summarizes the results of transient molecular dynamics simulations for GNROs with varying widths. For both pristine and edge-functionalized nanoribbons, $G$ initially decreases sharply with increasing width and subsequently increases as the width becomes larger. The GNRO edges, regardless of functionalization, are more effective in transferring heat to water than the basal plane. Therefore, increasing the ribbon width reduces the edge-to-surface ratio, leading to a lower overall $G$ and explaining the initial decrease. It is observed that as the width of pristine nanoribbon increases from 3.03 nm to 3.55 nm, a significant difference in the passivation rate of the nanoribbon emerges, as shown in the insets of Fig.~\ref{fig:G}c. For the 3.03 nm case, the nanoribbon exhibits higher initial passivation after relaxation, and has its dangling edge bonds rapidly passivated by hydrogen and hydroxyl groups from decomposing water molecules. In contrast, the 3.55 nm nanoribbon is less passivated after relaxation, and its dangling bonds remain unpassivated for a longer portion of the simulation. This phenomenon is explained by the increased stability of dangling bonds at the edge of nanoribbons with increasing width, and ultimately results in a decrease in thermal coupling with the aqueous medium.

For sufficiently wide nanoribbons, $G$ converges to nearly identical values for pristine and functionalized cases, reflecting the diminishing influence of edges on thermal transport within graphene, as discussed in Section~\ref{sec:k}. The subsequent increase in $G$ for GNROs wider than approximately 25~nm arises from the enhancement of in-plane $\kappa$, which mitigates local hotspot formation in the graphene layer. As revealed in Ref.~\cite{Cui2025Carbon}, such hotspots originate from the competition among multiple heat-transfer pathways from graphene to the surrounding water.

\section{Conclusion}

In this work, we conducted a comprehensive molecular dynamics study to elucidate the role of edge versus surface functionalization in governing thermal transport between GOs and water. Leveraging a deep neural network interatomic potential trained on extensive \emph{ab initio} data, we simultaneously quantified the in-plane thermal conductivity $\kappa$ of GOs and the GO-water interfacial thermal conductance $G$. Our results demonstrate that edge functionalization with hydroxyl groups provides a substantial enhancement of $G$ between GNRO and water, exceeding 100\% at low edge oxidation ratios, primarily due to improved wettability and stronger interfacial interactions with water molecules. Importantly, this enhancement saturates at relatively low functionalization levels, reflecting the finite packing density of water near the graphene edges and competition among parallel heat-transfer pathways. In contrast to surface functionalization, which can increase interfacial conductance by nearly an order of magnitude, edge functionalization preserves the high in-plane $\kappa$ of graphene by minimizing disruption to its basal-plane lattice and flexural phonon transport.

We further revealed a non-monotonic dependence of $G$ on nanoribbon width, governed by the interplay between edge-dominated heat transfer and in-plane heat spreading within graphene. Narrow ribbons benefit from efficient edge-mediated heat dissipation, while wider ribbons exhibit enhanced conductance due to reduced hotspot formation enabled by higher in-plane $\kappa$. Cross-sectional heat flux analysis directly confirms the critical role of edge-induced phonon scattering and localization in shaping thermal transport behavior in finite-width graphene structures.

Overall, this study establishes edge functionalization as a promising strategy for optimizing graphene--water thermal interfaces, offering a favorable balance between enhanced interfacial heat transfer and preserved intrinsic thermal transport properties. These insights provide fundamental guidance for the rational design of graphene-based thermal management systems in aqueous and bio-relevant environments, where both efficient heat dissipation and high in-plane $\kappa$ are essential.

\section*{Acknowledgments}
J.C., H.C. and Y.W. acknowledge the National Science Foundation EPSCoR Research Infrastructure Program (Award Number OIA-2033424). M.L. and L.C. acknowledge the National Science Foundation  (Award number OIA-2132224). J.C. acknowledges the partial support of the Nevada NASA Space Grant Consortium Graduate Research Opportunity Fellowship (Award number 80NSSC25M7094). The authors acknowledge the support of Research and Innovation and the Cyberinfrastructure Team in the Office of Information Technology at the University of Nevada, Reno, for facilitation and access to the Pronghorn High-Performance Computing Cluster.
\section*{Supplementary Material}
Supplementary material contains results that are not included in the main text but may help the readers understand the ideas presented in this manuscript.

\section*{Author Contributions}
\textbf{John Crosby:} Writing - original draft (lead), Methodology (lead), Formal analysis (lead), Software (equal), Conceptualization (equal), Writing - review and editing (equal)
\textbf{Haoran Cui:} Supervision (lead), Writing - original draft (lead), Writing - review and editing (equal), Software(equal), Formal analysis (equal), Conceptualization (equal), Methodology (equal)
\textbf{Mehrab Lotfpour:} Writing - original draft (equal), Writing - review and editing (equal), Methodology (equal), Conceptualization (equal), Software (equal)
\textbf{Yan Wang:} Supervision (lead), Funding aquisition (lead), Writing - original draft (equal) Writing - review and editing (equal), Software (equal), Formal analysis (equal), Concenptualization (equal)
\textbf{Lei Cao:} Supervision (lead), Funding aquisition (lead), Writing - original draft (equal), Writing - review and editing (equal), Software (equal), Formal Analysis (equal), Conceptualization (equal)

\section*{Data Availability Statements}
The data that support the findings of this study are available from the corresponding author upon reasonable request.

\bibliographystyle{elsarticle-num} 
\bibliography{references}






\end{document}


\begin{frontmatter}

\title{Supplementary Materials for ``Enhanced Graphene-Water Thermal Transport via Edge Functionalization without Compromising In-Plane Thermal Conductivity''}

\author[inst1]{John Crosby}
\author[inst1]{Haoran Cui\corref{cor1}}
\ead{hcui@unr.edu}

\author[inst1]{Mehrab Lotfpour}

\author[inst1]{Yan Wang\corref{cor2}}
\ead{yanwang@unr.edu}

\author[inst1]{Lei Cao\corref{cor3}}
\ead{leicao@unr.edu}

\cortext[cor1]{Corresponding author}
\cortext[cor2]{Corresponding author}
\cortext[cor3]{Corresponding author}

\affiliation[inst1]{organization={Department of Mechanical Engineering, University of Nevada, Reno},
            city={Reno},
            state={NV},
            postcode={89557},
            country={USA}}

\end{frontmatter}

\renewcommand{\thetable}{S\arabic{table}}
\renewcommand{\thefigure}{S\arabic{figure}}
\section{Convergence study for equilibrium molecular dynamics}
In equilibrium molecular dynamics simulations (EMD), the Green-Kubo formalism requires periodic boundary conditions in the direction along which thermal conductivity is evaluated. Consequently, the graphene nanoribbon oxide (GNRO) is effectively treated as infinite or "bulk" in this direction. Although extrinsic boundary scattering is absent in this configuration, the calculated $\kappa$ may still exhibit finite-size effects if the periodic length is insufficient to capture the full spectrum of contributing phonon modes. In particular, long-wavelength phonons can be artificially excluded when their mean free path exceeds the periodic dimension, resulting in an underprediction of $\kappa$. 

\begin{figure}[H]
    \centering
\includegraphics[width=0.8\textwidth, keepaspectratio]{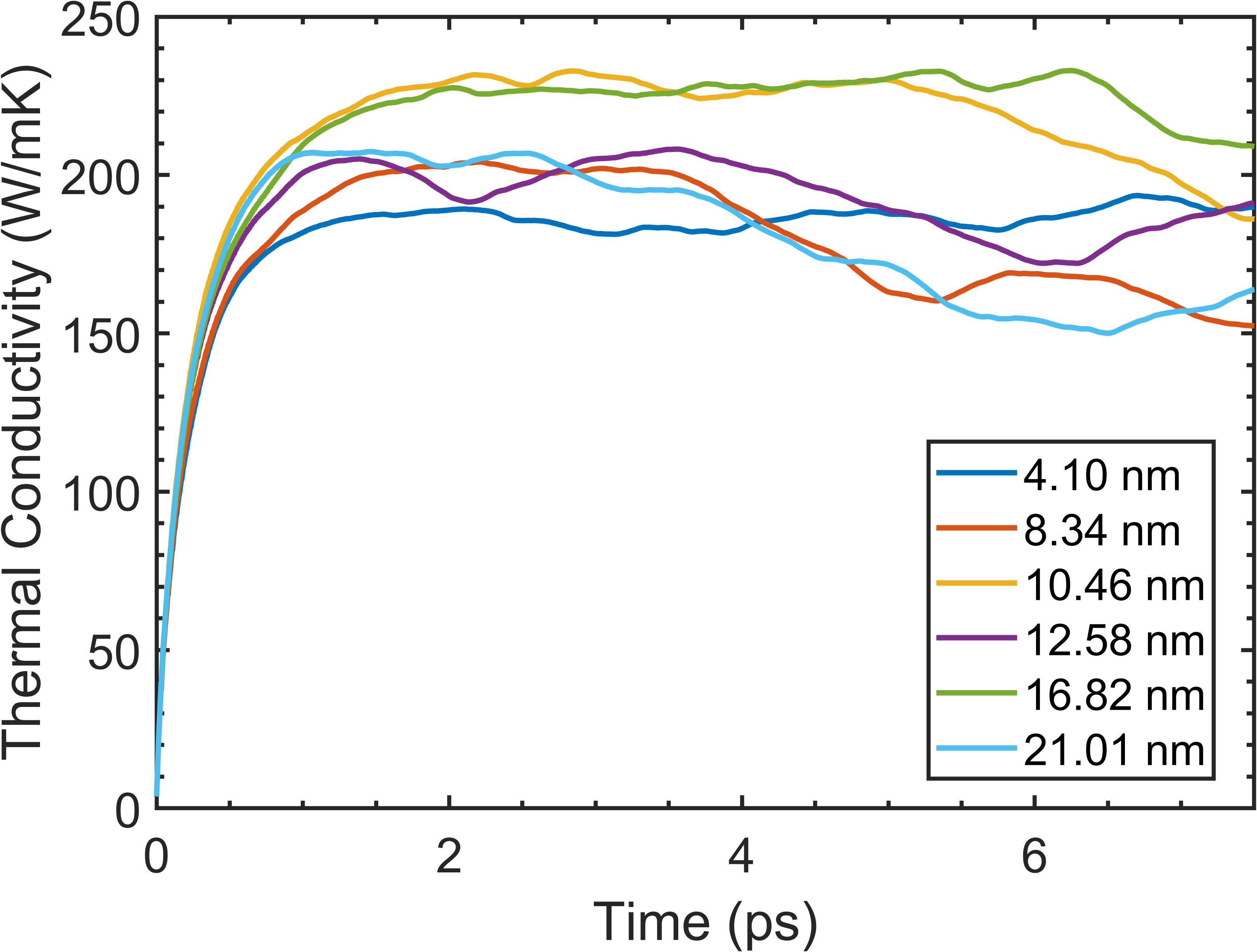}
    \caption{EMD convergence study results for varying length GNROs in the periodic direction.}
    \label{fig:supp_convergence}
\end{figure}

To ensure all relevant phonon contributions are captured, we performed a convergence study of $\kappa$ as a function of periodic-direction length for GNRO. Results of this convergence study are shown in Fig.~\ref{fig:supp_convergence}. Periodic-direction lengths in the range of 4.10 nm to 21.01 nm are simulated, and we observe $\kappa$ converges to approximately 200-225~W/m-K for lengths equal to or exceeding 8.34 nm. Beyond this length, further increases in the periodic length produce negligible changes in $\kappa$, indicating that all contributing phonon modes are sufficiently captured at this domain length. Based on this analysis, a periodic-direction length of 8.34 nm is selected for all GNRO simulations in this work, including both equilibrium and non-equilibrium approaches. 
\textbf{}

\section{Cross-sectional heat flux distribution for pristine graphene nanoribbon}

\begin{figure}[H]
    \centering
\includegraphics[width=0.8\textwidth, keepaspectratio]{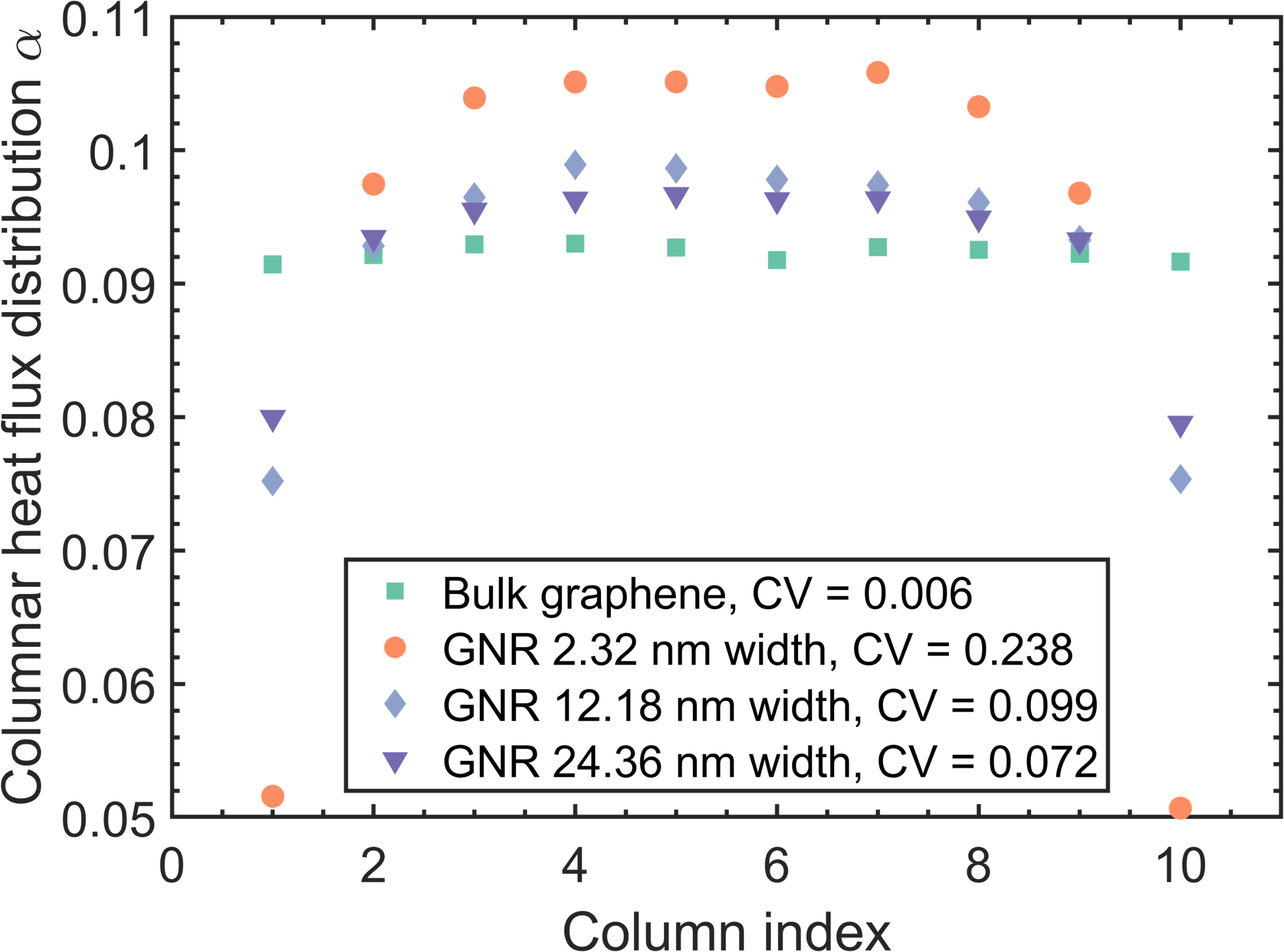}
    \caption{Cross-sectional heat flux distribution for pristine graphene}
    \label{fig:supp_CrossSectionalHeatFlux_Pristine}
\end{figure}

Figure \ref{fig:supp_CrossSectionalHeatFlux_Pristine} shows the cross-sectional columnar heat flux distribution ($\alpha$) for unfunctionalized graphene nanoribbon (GNR) of varying widths. The coefficient of variance (CV) is defined as the ratio of the standard deviation of $\alpha$ to its mean value, and quantifies the difference in heat transferred near the edges versus the center. The results indicate that edge localization and scattering effects are most pronounced for narrower GNRs: the 2.32 nm width GNR demonstrates the least heat transferred in the columns at the edges of the GNR, and consequently, the highest CV. The $\alpha$ profile flattens as GNR width increases, with more heat transferred in edge columns and lower CV values, approaching bulk graphene.

\textbf{}